\documentstyle[11pt,paspconf,epsf]{article}

\begin{document}

\title{Photometric Redshifts: A New Tool for Studying High-Redshift Clusters}
\author{L.M. Lubin \& R.J. Brunner}
\affil{Palomar Observatory, 105-24, California Institute of Technology, 
Pasadena, CA 91125}

\begin{abstract}

We present the first results of our application of photometric
redshifts to the study of galaxy populations in high-redshift
clusters.  For this survey, we are examining a sample of galaxy
clusters at $z > 0.6$ which have already been well-studied in the
optical and infrared wavelengths (Oke, Postman \& Lubin 1998). Our
main goal is to use photometric redshifts to delineate accurately
between field and cluster galaxies. Once we isolate the cluster
galaxies, we can directly study the properties of the galaxy
population in each high-redshift cluster. Specifically, we are
studying the cluster morphological fractions, the morphology-density
relation, and the large scale structure distribution.  Although we
have encountered some operational problems with our photometric
redshift technique, early results suggest that this procedure will
become a significant tool in studying high-redshift clusters.

\end{abstract}


\section{Introduction}

The techniques involved in calculating photometric redshifts have
become increasingly more sophisticated and precise. Accuracies of
$\sigma_{z} = 0.05$ can be routinely achieved (Brunner et al.\ 1999
and references therein). Therefore, it is now possible to use
photometric redshifts to answer specific scientific questions. In
light of this, we have begun a program to use accurate photometric
redshifts to study the galaxy populations in high-redshift clusters of
galaxies. For this study, we have examined galaxy clusters at $z >
0.6$ which are the subject of an extensive spectroscopic, photometric
and morphological survey by Oke, Postman \& Lubin (1998). Oke et al.\
(1998) have compiled an unprecedented amount of observational data for
each cluster, including deep $BVRIK$ photometry, over 150 high-quality
Keck spectra, and high-spatial-resolution WFPC2 imagery from HST.

Using the existing $BVRIK$ imaging and redshift data, we have refined
an empirical technique to measure photometric redshifts for all
galaxies down to $R = 24$ in each cluster field (see Brunner et al.\
1999). From our calibration sample, we are able to determine both a
redshift which is accurate to $\Delta \sigma \leq 0.05$ (Figure 1), as
well as an estimated redshift error for each galaxy in our catalog.
This approach works well for almost all types of galaxies over our
redshift range of interest. Specifically, we are able to measure
photometric redshifts for normal blue, star-forming galaxies and red,
elliptical-like galaxies which are expected to comprise the vast
majority of cluster members.

With these photometric redshifts, we can accurately delineate between
field and cluster galaxies, by utilizing the statistical nature of
photometric redshift estimation (Brunner 1997). In practice, this
reduces to selecting all galaxies which have a probability higher than
a predetermined threshold of lying within a redshift shell centered on
the cluster. The redshift probability distribution function (PDF) for
each galaxy is calculated assuming a Gaussian PDF with mean and sigma
given by the photometric redshift and redshift error estimates. Both
the probability threshold and the width of the redshift shell are free
parameters which we estimate using a maximum likelihood
technique. Consequently, we can study directly the galaxy populations
of the high-redshift clusters.  In these proceedings, we present the
first results of our work, as well as a discussion of the operational
difficulties that we have currently encountered.

\begin{figure}
\epsfysize=2.5in
\centerline{\epsfbox{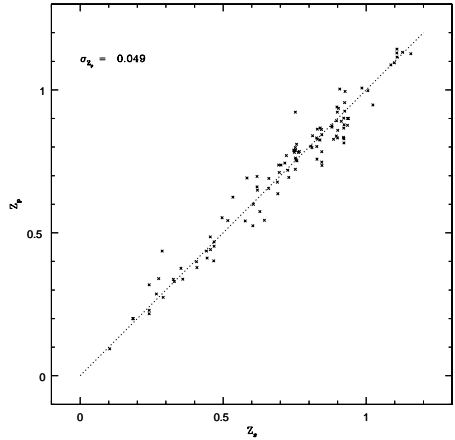}}
\caption{Spectroscopic redshift ($z_s$) versus photometric redshift
($z_p$) using $BRIK$ data from the original Oke, Postman \& Lubin
(1998) survey. Based on this data, it is possible to determine a
redshift to an accuracy of $\Delta \sigma \leq 0.05$.}
\end{figure}

\section{The Scientific Goals and The Need for Photometric Redshifts}

For this study, our principal scientific goals are (1) to study the
distribution of galaxy morphologies in high-redshift clusters; (2) to
measure the morphology-density relation at high redshift; and (3) to
map large scale structure at high redshift.  Traditionally, all of
these studies have been performed using a statistical analysis where
the contamination rate of the foreground and background galaxies had
to be estimated and subtracted from the cluster field. This technique
was necessary because of the limited number of spectroscopically
confirmed cluster members.  Unfortunately, considerable uncertainty
(and the resultant systematic effects) can be introduced if the
properties of the background population have not been estimated
correctly. Some issues which adversely affect this procedure are : (1)
the background contamination can be very large at high redshift (as
high as 85\%; see Oke et al.\ 1998); (2) the morphological mix of the
background population is normally derived from the pre-classified
field counts of the HDF and MDS (van den Bergh et al.\ 1996; Abraham
et al.\ 1996). Visual typing can be very subjective; therefore, using
classifications by other classifiers can introduce an additional
uncertainty (Naim et al.\ 1995a,b); and (3) there is a strong
variation in galaxy color in high-redshift clusters (including a
dominant population of red galaxies and an increased fraction of blue
galaxies). Consequently, we do not want to discount inadvertently any
part of the total cluster population.  With photometric redshifts, we
can directly examine the cluster population by choosing only those
galaxies which have a high probability of belonging to the cluster.

\section{The Preliminary Results}

In Figure 2, we present the F814W WFPC2 image of CL1324+3011 at $z =
0.76$.  This cluster is clearly defined by the central concentration
of bright, elliptical-like galaxies.  Based on the spectroscopic
survey of Oke et al.\ (1998), there are 18 galaxies which are
confirmed cluster members (Postman, Lubin \& Oke 1999). These galaxies
are circled in Figure 2.  Based on our photometric redshifts, we have
identified all those galaxies in the HST field-of-view which are
likely cluster members. These galaxies are indicated by diamonds.
From this figure, we can see that using photometric redshifts we have
done a reasonable job with our initial attempt at identifying all
those spectroscopically confirmed cluster members.  In addition, we
have identified over a factor of 2 more cluster members than the
spectroscopy alone. However, there are some confirmed cluster members
which were not identified by the photo $z$ method. Typically, these
galaxies have low luminosities and are compact. Many but not all have
have been visually classified as early-type galaxies, which we naively
expect based on our current requirement for accurate multi-band
photometry.

\begin{figure}
\plotone{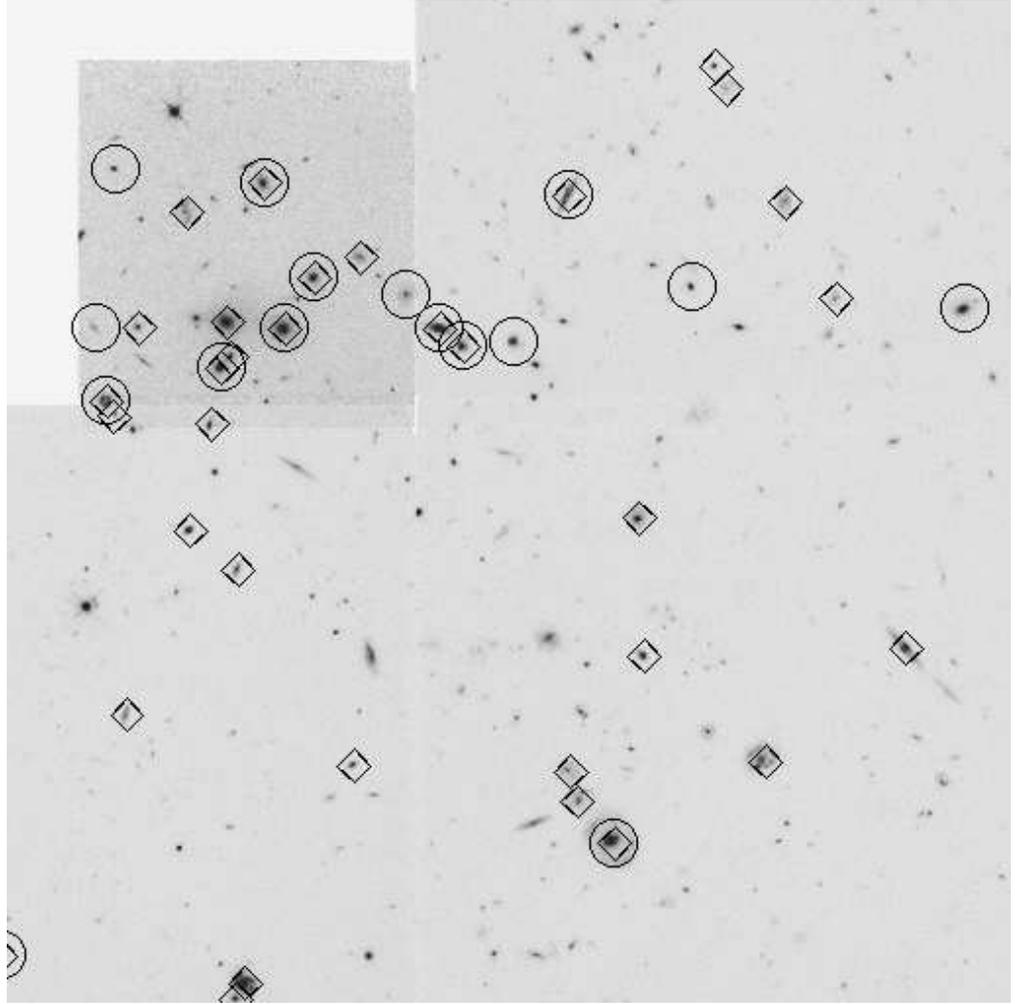}
\caption{The HST image of CL1324+3011 taken with the WPFC2 in the
F814W filter.  The cluster center is clearly defined by the
concentration of bright, elliptical-like galaxies. The
spectroscopically confirmed cluster members are indicated by circles,
while those galaxies believed to be cluster members based on their
photometric redshifts are indicated by diamonds. Using our initial
photometric redshift estimates, we have done a reasonable job at
identifying all of the spectroscopically confirmed cluster members.}
\end{figure}

In Figure 3, we present the distribution of galaxy morphologies in two
high-redshift clusters, CL0023+0423 at $z = 0.83$ and CL1604+4304 at
$z = 0.90$.  For each cluster, we show two morphological distributions
-- one determined from the traditional statistical approach (Lubin et
al.\ 1998; 1999) and one determined from photometric redshifts alone.
We see in the case of CL0023+0423 that the two distributions agree
well. This system is dominated by blue, late-type (spiral and
irregular/peculiar) galaxies.  For the case of CL1604+4304, the two
distributions differ significantly.  While the statistical analysis
indicates that this cluster is strongly dominated by early-type
(elliptical and S0) galaxies, the photo $z$ analysis indicates a
significantly larger population of late-type galaxies. This
discrepancy may result from the fact that we are systematically
missing some fainter early-type galaxies with the photo $z$ method
(see above); however, it may also indicate that the statistical
approach is flawed and that we have overestimated the fraction of
late-type galaxies in the background population. At this point, we are
still in the process of determining the true distribution of
morphologies in these clusters.

\begin{figure}
\epsfysize=3in
\centerline{\epsfbox{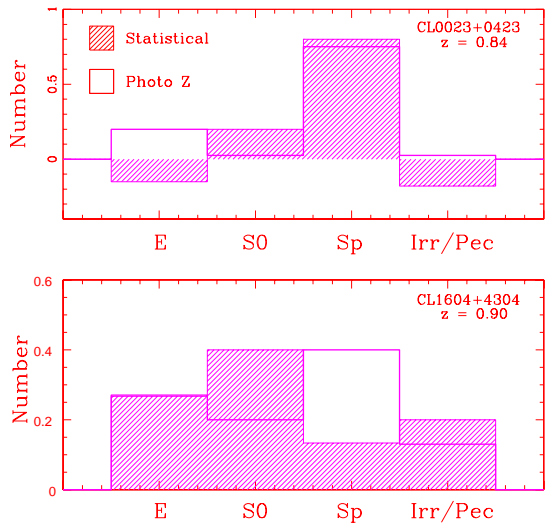}}
\caption{Distribution of galaxy morphologies in two high-redshift
clusters -- CL0023+0423 at $z = 0.84$ and CL1604+4304 at $z = 0.90$.
The shaded histograms represent the morphological distributions
calculated using the statistical method, while the solid-line
histogram represent those distributions calculated using the photo $z$
method.}
\end{figure}

\section{The Future}

Based on our work, we believe that photometric redshifts are an ideal
way to study the galaxy populations in high-redshift clusters.  This
method eliminates the need for extensive spectroscopic surveys and the
uncertainty of estimating the background contamination. We have,
however, encountered some operational problems.  These problems
include, firstly, large errors in the photometric redshifts when the
photometry is poor. Such errors limit our ability to gauge whether a
galaxy is a cluster member or not. Secondly, the lack of depth in the
bluest bands preferentially discriminates against early-type galaxies
and can, therefore, skew the cluster galaxy distribution.  Thirdly,
our photo $z$ method, as it now stands, cannot accurately estimate the
redshift of ``unusual'' galaxies, such as E+A galaxies which have a
spectrum characterized by strong Balmer absorption features and are
common in the cluster environment (e.g.\ Dressler \& Gunn 1983).
Consequently, these galaxies are currently excluded from the analysis.
We anticipate that we will be able to overcome these problems with
further refinement of the photometric redshift technique and a
specifically designed photometric survey which contains the
appropriate wavelength coverage and depth.  Therefore, we expect that,
in the near future, we will be able to utilize fully the capabilities
of this technique for cluster research.

\end{document}